\documentclass[conference]{IEEEtran}
\IEEEoverridecommandlockouts
\usepackage{cite}
\usepackage{amsmath,amssymb,amsfonts}
\usepackage{graphicx}
\usepackage{textcomp}
\usepackage{xcolor}

\usepackage{graphicx}
\usepackage{amsmath}
\usepackage{amssymb}
\usepackage{booktabs}
\usepackage{tabularray}
\usepackage{xcolor}
\usepackage{multirow}
\usepackage{multicol}
\usepackage{tabularx} 
\usepackage{pifont} 
\usepackage{verbatim} 
\usepackage{float} 
\usepackage[export]{adjustbox} 
\usepackage{siunitx}  
\usepackage{algpseudocode} 

\usepackage{subcaption}
\usepackage{tikz}
\usepackage{pgfplots}

\pgfplotsset{compat=newest}

\usetikzlibrary{arrows,arrows.meta, automata}
\usetikzlibrary{shapes.geometric,shapes.arrows,decorations.pathmorphing}
\usetikzlibrary{matrix,chains,scopes,positioning,arrows,fit}
\usetikzlibrary{patterns}
\usetikzlibrary{chains,backgrounds,calc}

\usetikzlibrary{shapes,arrows} 
\usetikzlibrary{calc} 

\definecolor{tu0}{rgb}{0.7451, 0.1176, 0.2353}

\definecolor{tu1}{rgb}{1.0000, 0.8039, 0.0000}
\definecolor{tu11}{rgb}{1.0000, 0.8627, 0.3020}
\definecolor{tu12}{rgb}{1.0000, 0.9020, 0.4980}
\definecolor{tu13}{rgb}{1.0000, 0.9412, 0.6980}
\definecolor{tu14}{rgb}{1.0000, 0.9608, 0.8000}

\definecolor{tu2}{rgb}{0.9804, 0.4314, 0.0000}
\definecolor{tu21}{rgb}{0.9882, 0.6039, 0.3020}
\definecolor{tu22}{rgb}{0.9882, 0.7137, 0.4980}
\definecolor{tu23}{rgb}{0.9922, 0.8275, 0.6980}
\definecolor{tu24}{rgb}{0.9961, 0.8863, 0.8000}

\definecolor{tu3}{rgb}{0.6902, 0.0000, 0.2745}
\definecolor{tu31}{rgb}{0.7529, 0.2000, 0.4196}
\definecolor{tu32}{rgb}{0.8431, 0.4980, 0.6353}
\definecolor{tu33}{rgb}{0.9216, 0.7490, 0.8196}
\definecolor{tu34}{rgb}{0.9529, 0.8510, 0.8902}

\definecolor{tu4}{rgb}{0.4863, 0.8039, 0.9020}
\definecolor{tu41}{rgb}{0.6431, 0.8627, 0.9333}
\definecolor{tu42}{rgb}{0.7412, 0.9020, 0.9490}
\definecolor{tu43}{rgb}{0.8431, 0.9412, 0.9686}
\definecolor{tu44}{rgb}{0.8980, 0.9608, 0.9804}

\definecolor{tu5}{rgb}{0.0000, 0.5020, 0.7059}
\definecolor{tu51}{rgb}{0.3020, 0.6510, 0.7961}
\definecolor{tu52}{rgb}{0.5490, 0.7765, 0.8667}
\definecolor{tu53}{rgb}{0.7490, 0.8745, 0.9255}
\definecolor{tu54}{rgb}{0.8510, 0.9255, 0.9569}

\definecolor{tu6}{rgb}{0.0000, 0.3255, 0.4549}
\definecolor{tu61}{rgb}{0.2510, 0.4941, 0.5922}
\definecolor{tu62}{rgb}{0.5490, 0.6941, 0.7529}
\definecolor{tu63}{rgb}{0.7490, 0.8314, 0.8627}
\definecolor{tu64}{rgb}{0.8510, 0.8980, 0.9176}

\definecolor{tu7}{rgb}{0.7765, 0.9333, 0.0000}
\definecolor{tu71}{rgb}{0.8431, 0.9529, 0.3020}
\definecolor{tu72}{rgb}{0.8863, 0.9647, 0.4980}
\definecolor{tu73}{rgb}{0.9333, 0.9804, 0.6980}
\definecolor{tu74}{rgb}{0.9569, 0.9882, 0.8000}

\definecolor{tu8}{rgb}{0.5373, 0.6431, 0.0000}
\definecolor{tu81}{rgb}{0.6784, 0.7490, 0.3020}
\definecolor{tu82}{rgb}{0.7686, 0.8196, 0.4980}
\definecolor{tu83}{rgb}{0.8588, 0.8941, 0.6980}
\definecolor{tu84}{rgb}{0.9059, 0.9294, 0.8000}

\definecolor{tu9}{rgb}{0.0000, 0.4431, 0.3373}
\definecolor{tu91}{rgb}{0.3020, 0.6118, 0.5373}
\definecolor{tu92}{rgb}{0.5490, 0.7490, 0.7020}
\definecolor{tu93}{rgb}{0.7490, 0.8588, 0.8353}
\definecolor{tu94}{rgb}{0.8549, 0.9176, 0.9059}

\definecolor{tu10}{rgb}{0.8000, 0.0000, 0.6000}
\definecolor{tu101}{rgb}{0.8706, 0.3490, 0.7412}
\definecolor{tu102}{rgb}{0.9216, 0.6000, 0.8392}
\definecolor{tu103}{rgb}{0.9608, 0.8000, 0.9216}
\definecolor{tu104}{rgb}{0.9804, 0.8980, 0.9608}

\definecolor{tu110}{rgb}{0.4627, 0.0000, 0.4627}
\definecolor{tu111}{rgb}{0.5961, 0.2510, 0.5961}
\definecolor{tu112}{rgb}{0.7294, 0.4980, 0.7294}
\definecolor{tu113}{rgb}{0.8392, 0.6980, 0.8392}
\definecolor{tu114}{rgb}{0.9216, 0.8510, 0.9216}

\definecolor{tu120}{rgb}{0.4627, 0.0000, 0.3294}
\definecolor{tu121}{rgb}{0.6118, 0.3020, 0.5333}
\definecolor{tu122}{rgb}{0.7569, 0.5490, 0.6980}
\definecolor{tu123}{rgb}{0.8667, 0.7490, 0.8314}
\definecolor{tu124}{rgb}{0.9216, 0.8510, 0.9020}

\definecolor{tu130}{rgb}{0.0314, 0.0314, 0.0314}
\definecolor{tu131}{rgb}{0.3725, 0.3725, 0.3725}
\definecolor{tu132}{rgb}{0.5882, 0.5882, 0.5882}
\definecolor{tu133}{rgb}{0.7529, 0.7529, 0.7529}
\definecolor{tu134}{rgb}{0.8667, 0.8667, 0.8667}
\newcommand{\putindex}[3]{\vtop{\hbox{\hspace{#3} $#1$}
            \hbox{\raise 6mm \hbox{$\scriptscriptstyle #2$}}}}

\newcommand{\gradx}[0]{\vtop{\hbox{\rm grad}
            \hbox{\raise 2.5mm \hbox{\rm \hspace{2mm} \footnotesize x}}}}

\newcommand{\grady}[0]{\vtop{\hbox{\rm grad}
            \hbox{\raise 2.5mm \hbox{\rm \hspace{2mm} \footnotesize y}}}}

\newcommand{\grad}[1]{\vtop{\hbox{\rm grad}
            \hbox{\raise 2.5mm \hbox{#1}}}}

\newcommand{\btb}{     \begin{tabbing}             }
\newcommand{\bte}{     \end{tabbing}               }

 \definecolor{c_green_1}{RGB}{37, 190, 118} 

\definecolor{c_red_1}{RGB}{190, 37, 74} 

\definecolor{c_orange_1}{RGB}{252, 173, 3} 

\definecolor{c_bic_deg_orange1}{RGB}{255, 179, 102} 

\def\BibTeX{{\rm B\kern-.05em{\sc i\kern-.025em b}\kern-.08em
    T\kern-.1667em\lower.7ex\hbox{E}\kern-.125emX}}
\begin{document}

\title{A Lightweight Image Super-Resolution Transformer\\ Trained on Low-Resolution Images Only}

\author{\IEEEauthorblockN{Björn Möller, Lucas Görnhardt, Tim Fingscheidt}
\IEEEauthorblockA{\textit{Technische Universität Braunschweig} \\
\textit{Institute for Communications Technology}\\
Braunschweig, Germany \\
\{bjoern.moeller, lucas.goernhardt, t.fingscheidt\}@tu-bs.de}
}

\maketitle

\begin{abstract}
Transformer architectures prominently lead single-image super-resolution (SISR) benchmarks, reconstructing high-resolution (HR) images from their low-resolution (LR) counterparts.
Their strong representative power, however, comes with a higher demand for training data compared to convolutional neural networks (CNNs). 
For many real-world SR applications, the availability of high-quality HR training images is not given, sparking interest in LR-only training methods.
The LR-only SISR benchmark mimics this condition by allowing only low-resolution (LR) images for model training.
For a 4x super-resolution, this effectively reduces the amount of available training data to 6.25\% of the HR image pixels, which puts the employment of a data-hungry transformer model into question.
In this work, we are the first to utilize a lightweight vision transformer model with LR-only training methods addressing the unsupervised SISR LR-only benchmark.
We adopt and configure a recent LR-only training method from microscopy image super-resolution to macroscopic real-world data, resulting in our multi-scale training method for bicubic degradation (MSTbic).
Furthermore, we compare it with \mbox{reference} \mbox{methods} and prove its effectiveness both for a transformer and a CNN model.
We evaluate on the classic SR benchmark datasets Set5, Set14, BSD100, Urban100, and Manga109, and show superior performance over state-of-the-art (so far: CNN-based) LR-only SISR methods.
The code is available on GitHub\footnote{https://github.com/ifnspaml/SuperResolutionMultiscaleTraining}.
\end{abstract}



\begin{figure}
    \centering
    \includegraphics[width=0.49\textwidth]{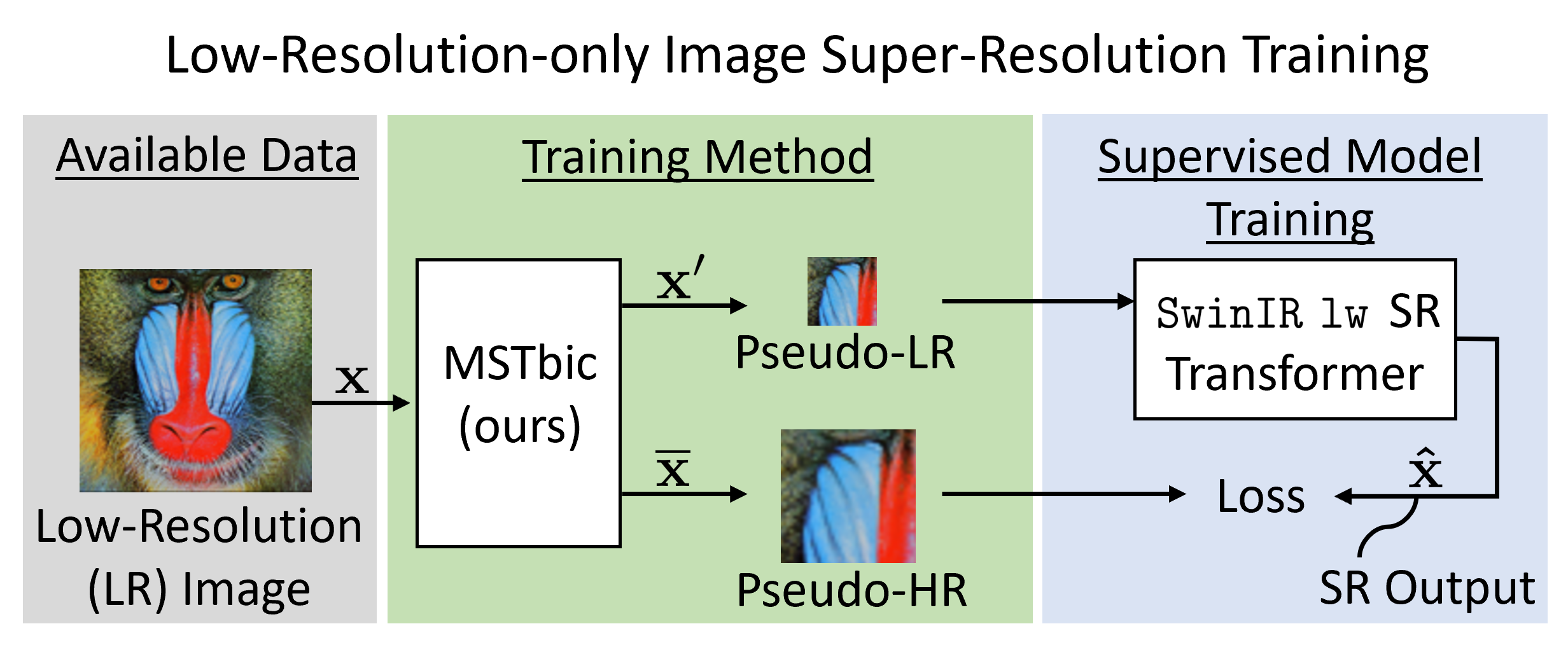}
    \caption{\textbf{Low-resolution-only training}: Overview over our proposed approach to train single-image super-resolution (SISR) models only with low-resolution (LR) data, consisting of our proposed multi-scale training method for bicubic degradation  (\textbf{MSTbic}). It creates pseudo-LR/HR training pairs enabling supervised \textbf{training of a \texttt{SwinIR} lightweight (lw) transformer model}. The LR-only SISR benchmark provides only low-resolution (LR) images during training, while high-resolution (HR) test data is used for evaluation.}
    \label{fig:page1_figure}
\end{figure}

\section{Introduction} \label{sec:intro}
Single-image super-resolution (SISR) models aim to construct a high-resolution (HR) output image from a corresponding low-resolution (LR) input image.
Conventionally, the SISR performance is measured in terms of image similarity between an SISR HR output image reconstructed from a degraded LR version and the original HR image. Also, the test images are taken from multiple benchmark datasets.
The established benchmarks such as SISR on DIV2K, Set5, Set14, BSD100, Urban100 and Manga109 were long led by convolutional neural networks (CNNs) \cite{ahn_fast_2018, zhang_residual_2018, ledig_photo-realistic_2017, tong_image_2017, lim_enhanced_2017, dong_learning_2014}, but were superseded quite recently by transformer models \cite{hsu_drct_2024,chen_activating_2023, chen_dual_2023, zhang_swinfir_2022, liang_swinir_2021}.

CNNs' main component are convolutional layers applying filters across the input to capture local spatial features. 
By progressively aggregating spatial information in deeper layers, hierarchies of features can be learned.
Transformers, originally designed for natural language processing \cite{vaswani_attention_2017}, excel at modeling long-range dependencies and global context by employing self-attention, which allows to process large image regions or even the entire image at once. 
The self-attention mechanism can model relationships between distant pixels or image features and dynamically weights their importance.
In contrast to CNNs, which rely on local convolutions and hierarchical feature extraction, transformers offer more flexibly and have a lower inductive bias induced by the architectural design.
We leverage these advantages and adopt a transformer-based lightweight \texttt{SwinIR} \cite{liang_swinir_2021} model in this work.

The absence of high-quality HR training images in many real-world SISR applications has led to the exploration of LR-only training methods.
For the LR-only task, training data is restricted to LR images, while evaluation is still conducted on HR images, defining an unsupervised task.
To still enable SR model training, some approaches \cite{shocher_zero-shot_2018, ahn_simusr_2020, moller_super-resolution_2023, moller_low-resolution-only_2024} generate pseudo LR/HR training pairs from the available LR images.
A recent method \cite{moller_super-resolution_2023} addresses microscopy images, while previous approaches \cite{shocher_zero-shot_2018, ahn_simusr_2020} for macroscopic images report on the LR-only SISR benchmark datasets and show interesting results that are almost competitive with supervised HR-trained models.
We adapt the LR-only multi-scale training method from microscopy \cite{moller_super-resolution_2023} to macroscopic images of the LR-only benchmark data under a bicubic degradation condition and consequently propose MSTbic, see Fig.\ \ref{fig:page1_figure}.

So far, only CNN-based models have been reported on the LR-only SISR benchmark, while CNNs had already been outperformed by transformers on other SISR benchmarks.
Still, it is not obvious whether this also applies to a transformer model under a limited data condition as present in the LR-only benchmark.
As transformers inherently make less assumptions about the data structure, they need to compensate for this low inductive bias by requiring extensive training data.
However, in the LR-only SISR benchmark, the amount of training information is reduced to 6.25\% of the original HR pixels for a 4x SR task.
In addition, training image quality is also lowered by the degradation function creating the LR images.

In this work, we first employ a transformer-based lightweight \texttt{SwinIR} model for LR-only SISR. Second, we adopt an LR-only training method from microscopy SISR to macroscopic data. Third, we compare it with reference methods and prove its effectiveness both for a transformer and a CNN model.
Finally, we set a new state of the art for the LR-only SISR benchmark.
\section{Related Work}\label{sec:rel_work}

\paragraph{CNNs and transformers for image super-resolution}
Deep learning-based SISR approaches primarily utilized CNNs due to their exceptional ability to extract local patterns. 
Initially featuring simple models with few layers \cite{dong_learning_2014}, improvements such as adding residual connections \cite{lim_enhanced_2017} or sub-pixel convolutions \cite{shi_real-time_2016} led to high-performance SR models\cite{ahn_fast_2018, zhang_residual_2018, ledig_photo-realistic_2017, tong_image_2017, lim_enhanced_2017, dong_learning_2014}.
Although effective, their performance is constrained by the locality of the feature extraction, imposed by the receptive field of the convolutional filters.
The essence of the transformer architecture \cite{vaswani_attention_2017} is the self-attention mechanism, which allows to model long-range dependencies and capture global relationships within the input sequence.
Initially developed for natural language processing \cite{vaswani_attention_2017} and later adopted to high-level vision tasks \cite{dosovitskiy_image_2021}, transformers since became predominant in super-resolution \cite{hsu_drct_2024,chen_activating_2023, chen_dual_2023, zhang_swinfir_2022, liang_swinir_2021}.
Compared to CNNs, Transformers have less inductive bias and therefore need more training data to realize their modeling potential \cite{chen_activating_2023}.
However, standard global self-attention scales quadratically with input size leading to high computational costs. 
To combat the computational demand, many approaches restrict the self-attention to local regions \cite{liang_swinir_2021, zhang_accurate_2022, chen_cross_2022, chen_activating_2023, hsu_drct_2024}.
\texttt{SwinIR} \cite{liang_swinir_2021} introduced local self-attention within shifting windows to still allow information exchange between neighboring local feature regions.
As the lightweight (lw) configuration provides an efficient trade-off between performance and computational cost, we deploy the \texttt{SwinIR lw} in our method.

\paragraph*{LR-only image super-resolution}
For some real-world SR applications high-quality, high-resolution ground-truth images cannot be acquired.
The LR-only image SR benchmark simulates this unsupervised condition, allowing only LR images during training while testing on HR images.
Training approaches \cite{shocher_zero-shot_2018, soh_meta-transfer_2020, ahn_simusr_2020, moller_super-resolution_2023} create pseudo-LR/HR training pairs to enable supervision for SR model training. 
Shocher et al.\ pioneered the task with a zero-shot super-resolution (ZSSR) \cite{shocher_zero-shot_2018} approach, that aims to utilize the internal image statistic of the test image by training a few-layer neural network for SR on the test image itself. For this, the LR test image was augmented by scaling it to multiple smaller resolutions before creating pseudo-training pairs. 
This online training, while flexible, has the disadvantage of long inference times.
Meta-transfer ZSSR (MZSR) \cite{soh_meta-transfer_2020} addresses this by adding pre-training and meta-transfer learning, reducing the training steps required at runtime. 
Following ZSSR's augmentation scheme, SimUSR \cite{ahn_simusr_2020} takes it to an offline training setting, allowing a dataset of multiple LR training images and leverages more sophisticated SR models. This drastically reduced inference time and improved the SR result.
SimUSR, however, was employed solely with CNNs, using an augmentation scheme that was purely dependent on downscaling the LR images.
Multi-scale training (MST) \cite{moller_super-resolution_2023} also includes image upscaling as data augmentation prior to creating the pseudo-LR/HR training pair.
It was, however, developed and deployed for atomic-scale microscopy images and uses large scaling factors of up to 250\%.
Some LR-only approaches \cite{wang_unsupervised_2021, neshatavar_icf-srsr_2023} address blind SISR tasks with unknown degradation, whereas the LR-only benchmark uses bicubic degradation.
In this work, we adapt MST to the macroscopic images of the LR-only benchmark data and leverage the additional upscaling augmentation to increase data variety and train the recent lightweight transformer-based \texttt{SwinIR lw} SR model.
\begin{figure*}
    \centering
            \centering
	      \input{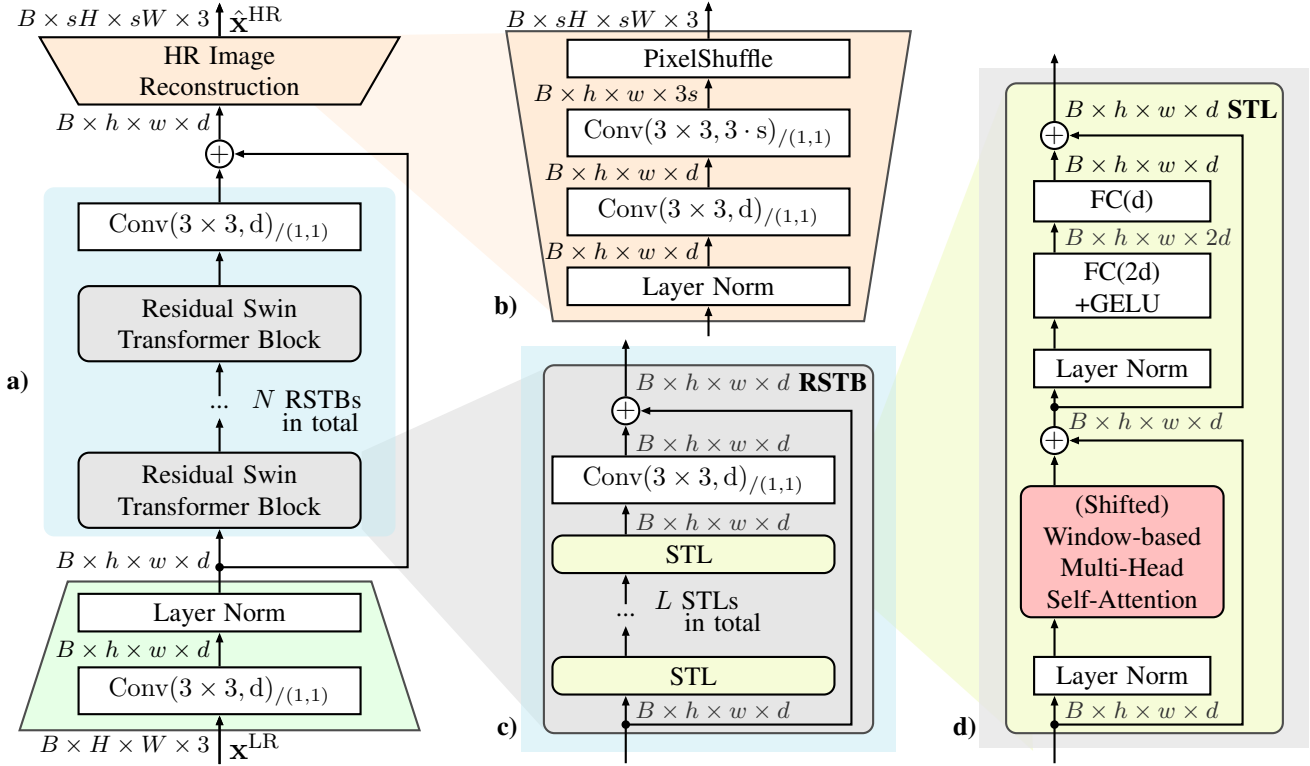}
            \caption{Architecture components of \texttt{SwinIR}\cite{liang_swinir_2021} for single-image super-resolution (SISR): (a) full architecture overview, (b) high-resolution image reconstruction module, (c) residual Swin transformer block (RSTB), (d) Swin transformer layer (STL).}
            \label{fig:SwinIR}
            \vspace*{-1mm}
\end{figure*}
\section{Method}\label{sec:method}
We first describe the \texttt{SwinIR lightweight} transformer model for SISR and then present our proposed LR-only training method MSTbic.

\subsection{Image Super-Resolution Transformer}
%
For the LR-only SISR task, we adapt a Swin image reconstruction (\texttt{SwinIR}) model \cite{liang_swinir_2021} introduced by \mbox{Liang et al.}, the model architecture of which is shown in \mbox{Fig.\ \ref{fig:SwinIR}a}.
It consists of a shallow feature extraction module, a deep feature extraction part and a HR image reconstruction module, the latter shown in \mbox{Fig.\ \ref{fig:SwinIR}b}.
The deep feature extraction is based on the structure of the \texttt{Swin} transformer \cite{liu_swin_2021} and utilizes $N$ sequential residual Swin transformer blocks (RSTBs), followed by a convolutional layer and residual connection.
Each RSTB implements $L$ consecutive and identical Swin transformer layers (STLs), also followed by a convolutional layer and jointly bypassed by a residual connection, as shown in Fig.\ \ref{fig:SwinIR}c. 
The STLs are built from the typical vision transformer block components \cite{dosovitskiy_image_2021}, which are an attention module, a multi-layer perceptron and normalization layers. Following \texttt{Swin}, a window-based multi-head self-attention module (W-MHSA) and two fully-connected layers (FCs) with a GELU activation \cite{hendrycks_gaussian_2023} are employed, each with a preceding layer norm and a subsequent residual connection, which is shown in \mbox{Fig.\ \ref{fig:SwinIR}d}.
%
The W-MHSA utilizes a localized self-attention, only calculating attention weights for image tokens inside a window.
The position of this window alternates between two layers, so every other layer employs a shifted W-MHSA.
This allows information to be transferred between the otherwise local attention windows.

We use the lightweight (lw) configuration for the \texttt{SwinIR} architecture with 0.89M parameters. It utilizes $N\!=\!4$ RSTBs, $L\!=\!6$ STLs, a feature embedding dimensionality $d\!=\!60$ and six attention heads with a window size of 8 and shift size \mbox{of 4.}

\subsection{LR-only Training Method}
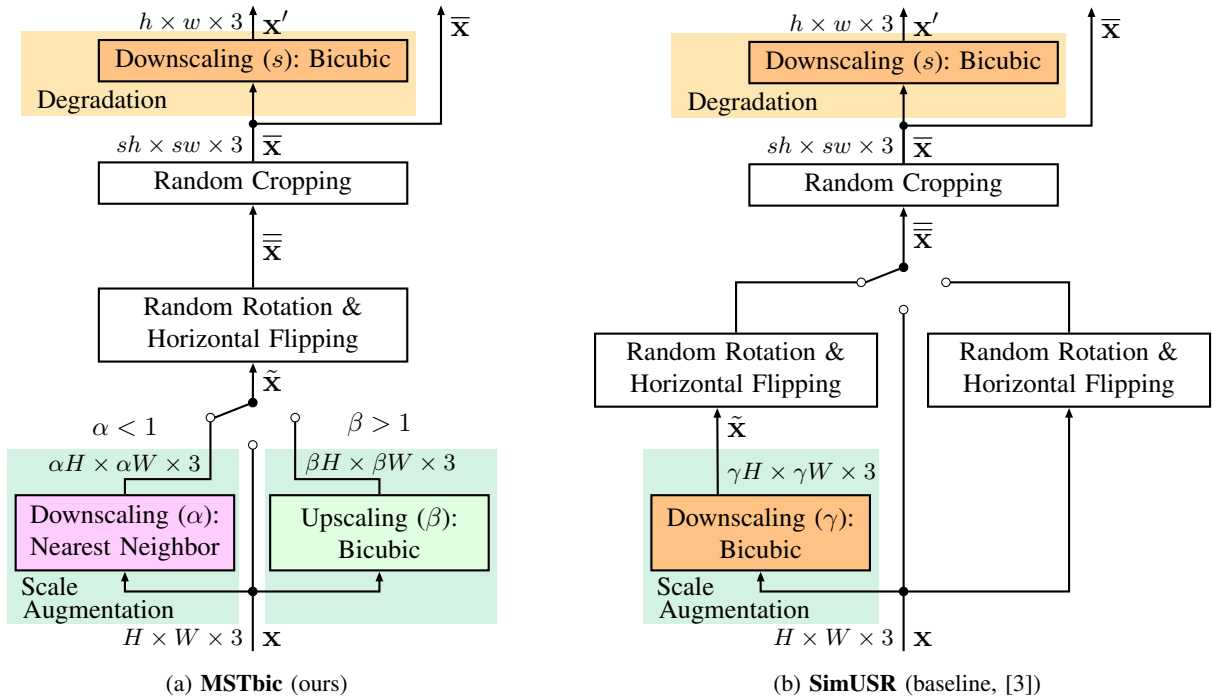
\begin{figure*}
    \centering
	\begin{subfigure}[b]{0.47\textwidth}
            \centering
	      \tikzset{
>=stealth',
punktchain/.style={
	rectangle,rounded corners, 
	draw=black, thick,
	text width=10em, 
	minimum height=0.3cm, 
	text centered,fill=white, 
	on chain=a},
line/.style={draw, thick, <-},
every join/.style={->, thick,shorten >=1pt},
}

\tikzset{
	>=stealth',
	punkt_mult/.style={
		circle, 
		inner sep=0.1mm, 
		draw=black, 
		text centered,fill=white, 
		on chain},
	line/.style={draw, thick, <-},
	every join/.style={->, thick,shorten >=1pt},
}

\tikzset{
	>=stealth',
	recblock/.style={
		rectangle, 
		draw=black, thick,
		text width=7.6em, 
		minimum height=0.5cm, 
		text centered,fill=white, 
		on chain=a},
	line/.style={draw, thick, <-},
	arrow/.style={draw, thick,->},
	loet/.style={draw,circle,fill=black,inner sep=0.3mm,outer sep=0.0mm},
	every join/.style={->, thick,shorten >=1pt},
    tipA/.tip = {Triangle[angle=40:4pt]},
	line/.style={draw, thick, <-},
	arrow/.style={draw, thick, -tipA},
}

\tikzset{
	>=stealth',
	trapez/.style={
		trapezium, 
		draw=black, thick,
		text width=10em, 
		minimum height=0.5cm, 
		text centered,fill=green!10, 
		on chain=a},
	line/.style={draw, thick, <-},
	arrow/.style={draw, thick,->},
	loet/.style={draw,circle,fill=black,inner sep=0.3mm,outer sep=0.0mm},
	every join/.style={->, thick,shorten >=1pt},
}

\begin{tikzpicture}[node distance=0.45cm,every node/.style={scale=1}] 

        \begin{scope}
	{  [start chain=a going above]

            \def\x{0} 
		\def\sc{1.1}
            \def\dimtextsize{\small}

            
            \node[draw=none, yshift=0cm] (dummy_start) {};
            
            \node[draw,circle,fill=black,inner sep=0.4mm,outer sep=0.0mm,yshift=-0.15cm, below=of dummy_start] (selector1_out) {};

            \node[draw=none,below=of dummy_start, outer sep=0mm,inner sep=0mm,yshift=-1cm] (input) {};

             \definecolor{light_magenta}{RGB}{255, 0, 255}
            \node[recblock, xshift=-1.7cm,minimum height=1cm,fill=light_magenta!20] (downscale) {Downscaling ($\alpha$):\\Nearest Neighbor};

            \definecolor{light_green}{RGB}{153, 255, 153}
            \node[recblock, right=of downscale, xshift=0cm, minimum height=1cm, fill=light_green!30] (upscale) {Upscaling ($\beta$):\\Bicubic}; 

             \node[draw,circle,fill=black,inner sep=0.4mm,outer sep=0.0mm,above=of selector1_out,xshift=0.0cm,yshift=1.95cm] (selector2_black) {};
            \node[draw,circle,fill=white,inner sep=0.4mm,outer sep=0.0mm, left=of selector2_black,xshift=0cm, yshift=-0.2cm] (selector2_white_left) {};
            \node[draw,circle,fill=white,inner sep=0.4mm,outer sep=0.0mm, right=of selector2_black,xshift=0cm, yshift=-0.2cm] (selector2_white_right) {};
           
            \node[draw,circle,fill=white,inner sep=0.4mm,outer sep=0.0mm, below=of selector2_black,xshift=0cm, yshift=0cm] (selector2_white_mid) {};

            \node[recblock, yshift=1.3cm, xshift=-1.68cm,text width=11em] (rotation) {Random Rotation \& \\ Horizontal Flipping};
            
            \node[recblock, yshift=0.65cm, xshift=0cm,text width=11em] (cropping) {Random Cropping};
            
            \node[recblock, yshift=0.6cm, text width=11em, fill=c_bic_deg_orange1!80] (degradation) {Downscaling ($s$): Bicubic};
                  
		\node[on chain=a,draw=none](out_node) {};
            \node[draw=none, right=of out_node, xshift=1.8cm](out_node2) {};
            
		\draw[ thick] (input) --node[left,text=black!100, yshift=-0.2cm]
		{\small $ H \times W \times 3$} node[right, text=black!100, yshift=-0.2cm]{\large $\mathbf{x}$} (selector1_out.south);

            \draw[-|,-tipA, thick,] ( selector1_out.west)node[left,text=black!100]{} -- ++(0.0,0.0) -|  (downscale.south);
		\draw[-|,-tipA, thick,] ( selector1_out.east)node[left,text=black!100]{} -- ++(0.0,0.0) -|  (upscale.south); 

            \draw[ thick,] ( downscale.north)node[left,text=black!100]{} --  ++(0.0,0.2) -|  node[left,text=black!100,yshift=0.2cm] {\dimtextsize $ \alpha H \times \alpha W \times 3$} (selector2_white_left.south);

            \draw[ thick,] ( upscale.north)node[left,text=black!100]{} -- ++(0.0,0.2) -| node[right,text=black!100,yshift=0.2cm] {\dimtextsize $ \beta H \times \beta W \times 3$} (selector2_white_right.south);

            \draw[thick] (selector2_black)  --node[left,text=black!100]
		{} (selector2_white_left.east);
  
            \draw[thick] (selector1_out.north)  --node[left,text=black!100]
		{} (selector2_white_mid.south);

            \draw[-|,-tipA, thick,] (selector2_black) --node[right,text=black!100]
		{\large $\tilde{\mathbf{x}}$} (rotation.south);

            \draw[-|,-tipA, thick,] (rotation.north) --node[right,text=black!100]
		{\large $\overline{\overline{\mathbf{x}}}$} (cropping.south);

             \draw[-|,-tipA, thick,] (cropping) --node[left,text=black!100, yshift=-0.3cm]
		{\dimtextsize $ sh \times  sw \times 3$} node[right,text=black!100,yshift=-0.3cm]
		{\large $\overline{\mathbf{x}}$} (degradation.south);
          
             \draw[-|,-tipA, thick,] (cropping.north)node[left,text=black!100]{} -- ++(0.0,0.5)node[loet]{} -| node[right,text=black!100,yshift=1.3cm]
		{\large $\overline{\mathbf{x}}$} (out_node2.south);
            
		\draw[-|,-tipA, thick,] ( degradation.north) --node[left,text=black!100] {\dimtextsize $  h \times w \times 3$} node[right,text=black!100] {\large $\mathbf{x}^\prime$} (out_node.south);
	}

    \begin{pgfonlayer}{background}
        \path (downscale.west |- downscale.north)+(-0.1,+0.6) node (g) {};
        \path (downscale.east |- downscale.south)+(+0.05,-0.7) node (h) {};
        \path[fill=c_green_1!20] (g) rectangle (h);
    \end{pgfonlayer}

    \begin{pgfonlayer}{background}
        \path (upscale.west |- upscale.north)+(-0.05,+0.6) node (g) {};
        \path (upscale.east |- upscale.south)+(+0.1,-0.7) node (h) {};
        \path[fill=c_green_1!20] (g) rectangle (h);
    \end{pgfonlayer}

     \node[below=of downscale,yshift=0.2cm, xshift=-0.35cm] (augmentation_tag) {Augmentation};
     \node[below=of downscale,yshift=0.5cm, xshift=-1cm] (scale_tag) {Scale};

     \node[above=of downscale,yshift=0.2cm, xshift=0cm] (alpha_tag) {$\alpha < 1$};

     \node[above=of upscale,yshift=0.2cm, xshift=0cm] (beta_tag) {$\beta > 1$};
    
    \begin{pgfonlayer}{background}
        \path (degradation.west |- degradation.north)+(-1.02,+0.1) node (g) {};
        \path (degradation.east |- degradation.south)+(+0.1,-0.45) node (h) {};
        \path[fill=c_orange_1!30] (g) rectangle (h);
    \end{pgfonlayer}

     \node[left=of degradation,yshift=-0.55cm, xshift=1.5cm] (degradation_tag) {Degradation};
    
    \end{scope}
    
 \end{tikzpicture}
            \caption{\textbf{MSTbic} (ours)}
        \end{subfigure}
        \begin{subfigure}[b]{0.47\textwidth}
		\centering
		\tikzset{
>=stealth',
punktchain/.style={
	rectangle,rounded corners, 
	draw=black, thick,
	text width=10em, 
	minimum height=0.3cm, 
	text centered,fill=white, 
	on chain=a},
line/.style={draw, thick, <-},
every join/.style={->, thick,shorten >=1pt},
}

\tikzset{
	>=stealth',
	punkt_mult/.style={
		circle, 
		inner sep=0.1mm, 
		draw=black, 
		text centered,fill=white, 
		on chain},
	line/.style={draw, thick, <-},
	every join/.style={->, thick,shorten >=1pt},
}

\tikzset{
	>=stealth',
	recblock/.style={
		rectangle, 
		draw=black, thick,
		text width=7.6em, 
		minimum height=0.5cm, 
		text centered,fill=white, 
		on chain=a},
	line/.style={draw, thick, <-},
	arrow/.style={draw, thick,->},
	loet/.style={draw,circle,fill=black,inner sep=0.3mm,outer sep=0.0mm},
	every join/.style={->, thick,shorten >=1pt},
    tipA/.tip = {Triangle[angle=40:4pt]},
	line/.style={draw, thick, <-},
	arrow/.style={draw, thick, -tipA},
}

\tikzset{
	>=stealth',
	trapez/.style={
		trapezium, 
		draw=black, thick,
		text width=10em, 
		minimum height=0.5cm, 
		text centered,fill=green!10, 
		on chain=a},
	line/.style={draw, thick, <-},
	arrow/.style={draw, thick,->},
	loet/.style={draw,circle,fill=black,inner sep=0.3mm,outer sep=0.0mm},
	every join/.style={->, thick,shorten >=1pt},
}

\begin{tikzpicture}[node distance=0.45cm,every node/.style={scale=1},  baseline=(current bounding box.center)] 

        \begin{scope}
	{  [start chain=a going above]

            \def\x{0} 
		\def\sc{1.1}
            \def\dimtextsize{\small}

            \node[draw=none, yshift=0cm] (dummy_start) {};
            
            \node[draw,circle,fill=black,inner sep=0.4mm,outer sep=0.0mm,yshift=-0.15cm, below=of dummy_start] (selector1_out) {};

            \node[draw=none,below=of dummy_start, outer sep=0mm,inner sep=0mm,yshift=-1cm] (input) {};

            \node[recblock, xshift=-1.9cm,minimum height=1cm,fill=c_bic_deg_orange1!80] (downscale) {Downscaling ($\gamma$):\\Bicubic};

            \node[recblock, yshift=0.7cm, xshift=4.1cm,text width=10em] (rotation_R) {Random Rotation \& \\ Horizontal Flipping};

             \node[recblock, yshift=0cm, xshift=-0.2cm,text width=10em, left=of rotation_R] (rotation_L) {Random Rotation \& \\ Horizontal Flipping};


             \node[draw,circle,fill=black,inner sep=0.4mm,outer sep=0.0mm,above=of selector1_out,xshift=0cm,yshift=3.75cm] (selector2_black) {};
            \node[draw,circle,fill=white,inner sep=0.4mm,outer sep=0.0mm, left=of selector2_black,xshift=0cm, yshift=-0.2cm] (selector2_white_left) {};
            \node[draw,circle,fill=white,inner sep=0.4mm,outer sep=0.0mm, right=of selector2_black,xshift=0cm, yshift=-0.2cm] (selector2_white_right) {};
             \node[draw,circle,fill=white,inner sep=0.4mm,outer sep=0.0mm, below=of selector2_black,xshift=0cm, yshift=0cm] (selector2_white_mid) {};

             \node[recblock, yshift=0.3cm, xshift=0cm,text width=11em, above=of selector2_black] (cropping) {Random Cropping};
            
            \node[recblock, yshift=0.6cm, text width=11em,fill=c_bic_deg_orange1!80] (degradation) {Downscaling ($s$): Bicubic};
                  
		\node[on chain=a,draw=none](out_node) {};
            \node[draw=none, right=of out_node, xshift=1.8cm](out_node2) {};
            
		\draw[ thick] (input) --node[left,text=black!100, yshift=-0.2cm]
		{\small $ H \times W \times 3$} node[right, text=black!100, yshift=-0.2cm]{\large $\mathbf{x}$} (selector1_out.south);

            \draw[thick] (selector1_out)  --node[left,text=black!100, yshift=0]
		{} (selector2_white_mid);
            \draw[-|,-tipA, thick,] ( selector1_out.east)node[left,text=black!100]{} -- ++(0.0,0.0) -|  (downscale.south);

		\draw[-tipA,thick] ( selector1_out.west)node[left,text=black!100]{} -- ++(0.0,0.0) -|   (rotation_R.south); 

            \draw[-tipA, thick] ( $(downscale.north) + (-0.57,0.0)$) node[right,text=black!100,yshift=0.9cm]{\large $\Tilde{\mathbf{x}}$} --   node[right,text=black!100,yshift=-0.3cm] {\dimtextsize $ \gamma H \times \gamma W \times 3$} ($(rotation_L.south) + (-0.25,0.0)$);

            \draw[thick] (selector2_black)  --node[left,text=black!100]
		{} (selector2_white_left.east);

            \draw[-tipA, thick] (selector2_black) --node[right,text=black!100]
		{\large $\overline{\overline{\mathbf{x}}}$} (cropping.south);
  
             \draw[thick](rotation_L.north) |-node[left,text=black!100, yshift=0.0cm]
		{} node[right,text=black!100,yshift=0cm]
		{} (selector2_white_left.west);

             \draw[thick]($(rotation_R.north) + (0.0,0.0)$) |- node[left,text=black!100, yshift=0.0cm] {}(selector2_white_right.east);
            
             \draw[-|,-tipA, thick,] (cropping) --node[left,text=black!100, yshift=-0.3cm]
		{\dimtextsize $ sh \times  sw \times 3$} node[right,text=black!100,yshift=-0.3cm]
		{\large $\overline{\mathbf{x}}$} (degradation.south);
          
             \draw[-|,-tipA, thick,] (cropping.north)node[left,text=black!100]{} -- ++(0.0,0.5)node[loet]{} -| node[right,text=black!100,yshift=1.3cm]
		{\large $\overline{\mathbf{x}}$} (out_node2.south);
            
		\draw[-|,-tipA, thick,] ( degradation.north) --node[left,text=black!100] {\dimtextsize $  h \times w \times 3$} node[right,text=black!100] {\large $\mathbf{x}^\prime$} (out_node.south);
	}

    \begin{pgfonlayer}{background}
        \path (downscale.west |- downscale.north)+(-0.1,+0.6) node (g) {};
        \path (downscale.east |- downscale.south)+(+0.1,-0.7) node (h) {};
        \path[fill=c_green_1!20] (g) rectangle (h);
    \end{pgfonlayer}

     \node[below=of downscale,yshift=0.2cm, xshift=-0.35cm] (augmentation_tag) {Augmentation};
     \node[below=of downscale,yshift=0.5cm, xshift=-1cm] (scale_tag) {Scale};
    
    \begin{pgfonlayer}{background}
        \path (degradation.west |- degradation.north)+(-1.02,+0.1) node (g) {};
        \path (degradation.east |- degradation.south)+(+0.1,-0.45) node (h) {};
        \path[fill=c_orange_1!30] (g) rectangle (h);
    \end{pgfonlayer}

     \node[left=of degradation,yshift=-0.55cm, xshift=1.5cm] (degradation_tag) {Degradation};
    
    \end{scope}
    
 \end{tikzpicture}  
		\caption{\textbf{SimUSR} (baseline,\cite{ahn_simusr_2020})}
		\label{fig:ogmst}
	\end{subfigure}
    \caption{\textbf{LR-only training methods}: (a) Our \textbf{proposed multi-scale training} with bicubic upscaling and bicubic degradation \textbf{(MSTbic)}, (b)~\textbf{SimUSR} \cite{ahn_simusr_2020} (\textbf{baseline} method). MSTbic leverages initial downscaling (nearest neighbor) and upscaling (bicubic), while SimUSR only uses downscaling (bicubic) for image scale augmentation.}
    \label{fig:LRonlyMethods}
    \vspace*{-1mm}
\end{figure*}
Our proposed LR-only multi-scale training method for bicubic degradation (MSTbic) creates pseudo-LR/HR training pairs from an LR training image leveraging  downscaling and upscaling for image augmentation. 
Input to the method is an RGB image $\mathbf{x} \in \mathbb{G}^{H \times W \times C}$, with the set of gray values $\mathbb{G}=\{0,1,...,255\}$ and with the height $H$, width $W$ and the number of color channels $C=3$.
The method's output are the paired pseudo-HR target image $\overline{\mathbf{x}}$ and pseudo-LR input image $\mathbf{x}^\prime$, as shown in Fig.\ \ref{fig:page1_figure} and in Fig.\ \ref{fig:LRonlyMethods}a. 

The first step of our LR-only training method is scale augmentation.
In this process, the LR training image $\mathbf{x}$ is either downscaled, or left at its original size, or upscaled, with respective probabilities of $44.4\overline{4}\%$, $11.1\overline{1}\%$, and $44.4\overline{4}\%$.
For \textit{downscaling, we deploy nearest neighbor interpolation} \cite{Burger2009}, which assigns each pixel in the downscaled image the value of the closest pixel in the original image, preserving the original colors. 
In case of \mbox{\textit{upscaling}}, we use bicubic interpolation \cite{Burger2009}, which calculates each pixel's value in the upscaled image based on the weighted average of the 16 nearest pixels from the original image, smoothing the resulting image.
The corresponding downscaling and upscaling factors $\alpha$, $\beta$ are chosen at random from predefined sets of four equidistant values between $\alpha^\mathrm{min}=0.9$ and $1.0$ for downscaling, or between $1.0$ and $\beta^\mathrm{max}=1.1$ for upscaling.
To handle special cases of particularly small LR images, only $\alpha$ values that result in an image size allowing for the subsequent cropping of pseudo-targets are used for downscaling.
Eventually, this yields a scale-augmented image $\tilde{\mathbf{x}}$, that has been resized in the range of ($\alpha^\mathrm{min}$, $\beta^\mathrm{max}) \!=\! (0.9,1.1)$ of the original image size.
Next, image $\tilde{\mathbf{x}}$ is randomly rotated by 0°, 90°, 180°or 270°. Combined with random horizontal flipping, this augmentation step leads to a total of eight possible image orientations for the resulting image $\overline{\overline{\mathbf{x}}}$.
To create the pseudo-HR training target $\overline{\mathbf{x}}$, a random section of size $sh \times sw \times 3$ is cropped from $\overline{\overline{\mathbf{x}}}$. 
Finally, this pseudo-HR target $\overline{\mathbf{x}}$ is degraded through bicubic downscaling by the super-resolution factor $s$ to create the corresponding pseudo-LR input $\mathbf{x}^\prime$. With this pseudo-LR/HR training pair ($\mathbf{x}^\prime$, $\overline{\mathbf{x}}$), an SR model can be trained in a supervised fashion.

The proposed method has been inspired by a multi-scale training (MST) method developed for grayscale microscopy images \cite{moller_super-resolution_2023}. In contrast to our version, MST randomly chooses the degradation function from a set of interpolation kernels, consisting of nearest neighbor\cite{Burger2009}, Lanczos\cite{Duchon1979}, bilinear\cite{Burger2009}, bicubic\cite{Burger2009}, box\cite{Gonzales2008}, and Hamming\cite{Hamming1983} interpolation, which renders the original MST method quite complex.
We, instead, use only bicubic downsampling to generate the final pseudo LR/HR-pair.
Also, the image is significantly more augmented in size using scaling factors of $\alpha^\mathrm{min}=0.25$ and $\beta^\mathrm{max}=2.5$ in MST's main configuration for the microscopy application.
In this work, we found that large scaling factors are rather unfavorable for macroscopic images showing a higher level of detail, so we deploy $\alpha^\mathrm{min}=0.9$ and $\beta^\mathrm{max}=1.1$, which we later justify with ablations.
\begin{table*}[!htb]
	\centering
	\caption{\textbf{LR-only SISR benchmark} with models {\bf trained} based {\bf on LR images of DF2K\textsuperscript{train}}: Method comparison on the {\bf 4x bicubic SR task} and benchmark datasets. We report PSNR (dB) and SSIM on the Y-channel. We fine-tuned our \texttt{SwinIR lw} model from 2x bicubic SR pre-trained weights. $\dagger$ marks results taken from literature. Best results are in bold font, second best underlined. Supervised SR results in the bottom table segment are given as a soft upper bound for reference only.}
   \resizebox{1\textwidth}{!}{
	\begin{tabular}{l|c|c@{\hskip 4pt}c|c@{\hskip 4pt}cc@{\hskip 4pt}cc@{\hskip 4pt}cc@{\hskip 4pt}cc@{\hskip 4pt}c}
		\toprule
		Method & \# of &\multicolumn{2}{c|}{DIV2K\textsuperscript{val}}& \multicolumn{2}{c}{Set5} 	& \multicolumn{2}{c}{Set14} & \multicolumn{2}{c}{BSD100} & \multicolumn{2}{c}{Urban100} &\multicolumn{2}{c}{Manga109} \\
             \;\;+{\tt model} & parameters & PSNR & SSIM & PSNR & SSIM & PSNR & SSIM & PSNR & SSIM & PSNR & SSIM & PSNR & SSIM \\ \midrule
		Bicubic & - & &                         & 28.44         & 0.8110         & 25.87         & 0.7056  & 25.98        & 0.6698         & 23.14 & 0.6592 & 24.93 & 0.7906\\
            \midrule
             ZSSR \cite{ahn_simusr_2020}{$\dagger$}       & 0.23M &-&-& 31.13         & 0.8796         & 28.01         & 0.7651 & 27.12         & 0.7211         & 24.61 & 0.7282 & 27.84 & 0.8657\\
            SimUSR \cite{ahn_simusr_2020} \\
		\;\;+{\tt CARN} \cite{ahn_simusr_2020}{$ \dagger $} & 1.14M &-&-& 31.94         & 0.8908         & 28.44         & 0.7786 & 27.49         & 0.7324         & 25.70 & 0.7740 & 30.03 & 0.9014\\
            MSTbic (ours) &&& \\
            \; +{\tt SwinIR lw} &0.89M &\textbf{30.66} & \textbf{0.8437} &32.30 &\textbf{0.8966} & 28.57 & \textbf{0.7884} &\textbf{27.73} & \textbf{0.7422} & \underline{26.38} & \textbf{0.7988} & \textbf{31.17} & \textbf{0.9160}\\
            \midrule
            %
            SimUSR \cite{ahn_simusr_2020} \\
            \;\;+{\tt RCAN} \cite{ahn_simusr_2020}{$\dagger$} & 15.6M &-&-&\textbf{32.40} &\underline{0.8962} &\textbf{28.71} & \underline{0.7860} & \underline{27.68}  & \underline{0.7394}         &\textbf{26.45} &\underline{0.7986} & \underline{30.73} & \underline{0.9124}\\ 
            \;\;+{\tt EDSR} \cite{ahn_simusr_2020}{$\dagger$} & 42.2M  &-&-& \underline{32.37}         & 0.8955         & \underline{28.70}         & 0.7855 & 27.66         & 0.7389         & 26.31 & 0.7940 & 30.59 & 0.9107\\ 
            \midrule
            \midrule
            Supervised SR\\
            \;\;{\tt SwinIR} \cite{liang_swinir_2021}{$\dagger$}&11.8M&-&-&32.92          & 0.9044         & 29.09        & 0.7950 & 27.92          & 0.7489       & 27.45 & 0.8254 & 32.03 & 0.9260\\
            \;\;{\tt SwinIR lw} \cite{liang_swinir_2021}{$\dagger$} &0.89M &-&-&32.44 & 0.8976         & 28.77        & 0.7858 & 27.69          & 0.7406       & 26.47 & 0.7980  & 30.92 & 0.9151\\ 
	  	\bottomrule
	\end{tabular}
    }
	\label{tab:LRonlyBenchmark}
\end{table*}

Fig.\ \ref{fig:LRonlyMethods}b shows the SimUSR \cite{ahn_simusr_2020} baseline. 
 It also utilizes scale augmentation, but in contrast to MSTbic, only downscaling is applied to the LR image $\mathbf{x}$. In addition, unlike our method, they use the bicubic kernel for downscaling to generate pseudo-HR images.
In SimUSR, the downscaling factor $\gamma\!\sim\!\mathcal{U}(\gamma^\mathrm{min},1)$ is randomly drawn from a uniform distribution within a lower bound of $\gamma^\mathrm{min}\!=\!0.5$ up to $1.0$.
Next, the downscaled image $\Tilde{\mathbf{x}}$ is also randomly rotated and randomly flipped.
SimUSR applies downscaling with subsequent random rotation and flipping with 50\% probability, or only random rotation and random flipping with 20\% probability, or via a bypass of both augmentations with a probability of 30\%.
\section{Experiments and Discussion}\label{sec:experiments}
\subsection{Experimental Setup}

\paragraph*{Datasets and metrics}
 Following the LR-only SISR benchmark, we use the LR images of DF2K (DIV2K train \cite{agustsson_ntire_2017} plus Flicker2K \cite{timofte_ntire_2017}, in total 3450 images) for model training (DF2K\textsuperscript{train}).
 We validate and optimize our method on the DIV2K validation set (DIV2K\textsuperscript{val}) \cite{agustsson_ntire_2017}. 
 For evaluation, we report results on the benchmark test datasets Set5 \cite{bevilacqua_low-complexity_2012}, Set14 \cite{zeyde_single_2010}, BSD100\cite{martin_database_2001}, Urban100 \cite{huang_single_2015} and Manga109 \cite{matsui_sketch-based_2017}.
LR images are created by bicubic downscaling.
We report PSNR (dB) \cite{Salomon2004} and SSIM \cite{wang_image_2004} on the Y-channel.
\paragraph*{Baselines} 
We benchmark the \texttt{SwinIR lw} transformer model against the Cascading Residual Network (\texttt{CARN})\cite{ahn_fast_2018}, which is a lightweight CNN model for SISR.
\texttt{CARN}'s architecture is build from ResNet blocks \cite{he_deep_2016} and introduces a cascading mechanism, which leverages local and global skip connections between blocks to propagate information from shallow to deeper layers.

For LR-only training methods, we compare with SimUSR\cite{ahn_simusr_2020} (see Fig.\ \ref{fig:LRonlyMethods}b) and MST \cite{moller_super-resolution_2023} as main baselines, which also follow the approach of creating pseudo-LR/HR training pairs based on the LR images.
Ahn et al.\cite{ahn_simusr_2020} use SimUSR + \texttt{CARN} for their lightweight model experiments achieving the so-far state of the art, so we pick it as the most important baseline in our investigations.
As a non-trainable baseline, we additionally report bicubic interpolation.
For further reference, we also show supervised \texttt{SwinIR lw} results from literature as an upper bound, keeping in mind that they use HR images for training.

\paragraph*{Training details}
The models were trained for 500k iterations, using the Adam optimizer \cite{kingma_adam_2015} with an initial learning rate of 0.002 and step-wise learning rate scheduling, following Liang et al. \cite{liang_swinir_2021}. We use a training target crop size of 256$\times$256 pixels with a batch size of 16. 
Our most successful model for the LR-only benchmark utilizes pre-trained weights from the 2x bicubic SR task, also based the low-resolution training images of DIV2K\textsuperscript{train}.
All other models for comparison or ablations were trained from scratch.
Models were trained on an \texttt{NVidia GTX 1080 Ti} using the \texttt{PyTorch} framework \cite{Paszke2019}.
Image scaling functions from the \texttt{Python Pillow Library} \cite{Clark2015} were used.

\subsection{Results and Discussion}
\begin{table*}[!htb]
	\centering
	\caption{\textbf{LR-only SISR benchmark results} for a \textbf{CNN} (\texttt{CARN}) and a \textbf{transformer} (\texttt{SwinIR lw}) with LR-only training methods \textbf{SimUSR} \cite{ahn_simusr_2020} and \textbf{MSTbic} (ours) on the {\bf 4x bicubic SR task}.
 We report PSNR (dB) and SSIM on the Y-channel with \textbf{standard deviations} in ($10^{-2}$) calculated over 3 runs. Models were {\bf trained from scratch} based {\bf on} {\bf LR images of DF2K\textsuperscript{train}}. Best results are in bold font, second best underlined.}
   \resizebox{\textwidth}{!}{
	\begin{tabular}{l@{\hskip 2pt}|@{\hskip 3pt}c@{\hskip 4pt}c@{\hskip 7pt}c@{\hskip 4pt}c@{\hskip 7pt}c@{\hskip 4pt}c@{\hskip 7pt}c@{\hskip 4pt}c@{\hskip 7pt}c@{\hskip 4pt}c@{\hskip 1pt}}
		\toprule
		{\tt Model} & \multicolumn{2}{c}{Set5} 	& \multicolumn{2}{c}{Set14} & \multicolumn{2}{c}{BSD100} & \multicolumn{2}{c}{Urban100} &\multicolumn{2}{c}{Manga109} \\
             + method  & PSNR & SSIM & PSNR & SSIM & PSNR & SSIM & PSNR & SSIM & PSNR & SSIM \\ \midrule
		Bicubic & 28.44         & 0.8110         & 25.87         & 0.7056  & 25.98        & 0.6698         & 23.14 & 0.6592 & 24.93 & 0.7906\\ 
        \midrule
            {\tt CARN} \\
            \;\;+SimUSR \cite{ahn_simusr_2020} & 31.84\textpm1 &  0.8919\textpm0.03   & 28.24\textpm2 & 0.7823\textpm0.06 & 27.53\textpm2 & 0.7350\textpm0.02 & 25.89\textpm2 & 0.7813\textpm0.13 & 30.26\textpm17 & 0.9067\textpm0.09 \\
            \;\;+MSTbic (ours) & 31.90\textpm2 & 0.8926\textpm0.04 & 28.25\textpm3 & 0.7825\textpm0.04 & 27.55\textpm2 & 0.7360\textpm0.05 & 25.88\textpm3 & 0.7803\textpm0.14 & 30.41\textpm5 & 0.9072\textpm0.05\\
            {\tt SwinIR lw}\cite{liang_swinir_2021} \\
            \;\;+SimUSR \cite{ahn_simusr_2020} & \underline{32.21}\textpm5 & \underline{0.8957}\textpm0.05 & \underline{28.45}\textpm4 & \underline{0.7865}\textpm0.06 & \underline{27.66}\textpm1 & \underline{0.7402}\textpm0.06 & \underline{26.31}\textpm4 & \underline{0.7949}\textpm0.17 & \underline{30.59}\textpm2 & \underline{0.9126}\textpm0.02\\
            \;\;+MSTbic (ours) & \textbf{32.25}\textpm1 & \textbf{0.8962}\textpm0.00 & \textbf{28.53}\textpm1 & \textbf{0.7883}\textpm0.01 & \textbf{27.71}\textpm0 & \textbf{0.7416}\textpm0.01 & \textbf{26.36}\textpm 1  & \textbf{0.7967}\textpm0.00 & \textbf{31.02}\textpm1  & \textbf{0.9152}\textpm0.01\\
	  \bottomrule
	\end{tabular}
        }
	\label{tab:method_comparison}
\end{table*}

\paragraph*{LR-only SISR benchmark}
We evaluate our method on the LR-only SISR benchmark comparing it with reference methods from literature in Table \ref{tab:LRonlyBenchmark}.
\textit{Within the lightweight methods' table segment} (second from top, right after bicubic), \textit{our approach significantly outperforms the baselines and sets a new state of the art across all test datasets}.
Our method improves over the earlier state-of-the-art method SimUSR + \texttt{CARN} with PSNR gains of 0.13 dB (Set14) up to 1.14 dB (Manga109), and SSIM improvements from +0.0058 (0.8908 vs.\ 0.8966) on Set5 up to +0.0248 (0.7740 vs.\ 0.7988) on Urban100.
Also, we reduce the performance gap to methods using larger CNN models (third table segment), even exceeding those in SSIM and on BSD100 and Manga109 on both metrics.
As can be seen in the bottom table segment, even for reference results from supervised \texttt{SwinIR lw} training, which are out of competition and separate from the benchmark, our LR-only training \mbox{MSTbic} surpasses those on BSD100 and Manga109 and most SSIM values.

To assess the individual effects of the transformer \mbox{architecture} and our LR-only training method MSTbic, we provide a cross comparison to the strongest baseline in \mbox{Table \ref{tab:method_comparison}}. We report the mean and standard deviations of three runs each and demonstrate a superior performance of \texttt{SwinIR lw} (0.89M parameters) over \texttt{CARN} (1.14M) in all metrics (PSNR and SSIM), as well as the  performance improvement from MSTbic training compared to SimUSR training, especially for the \texttt{SwinIR lw}. 
\begin{figure*}[h!]
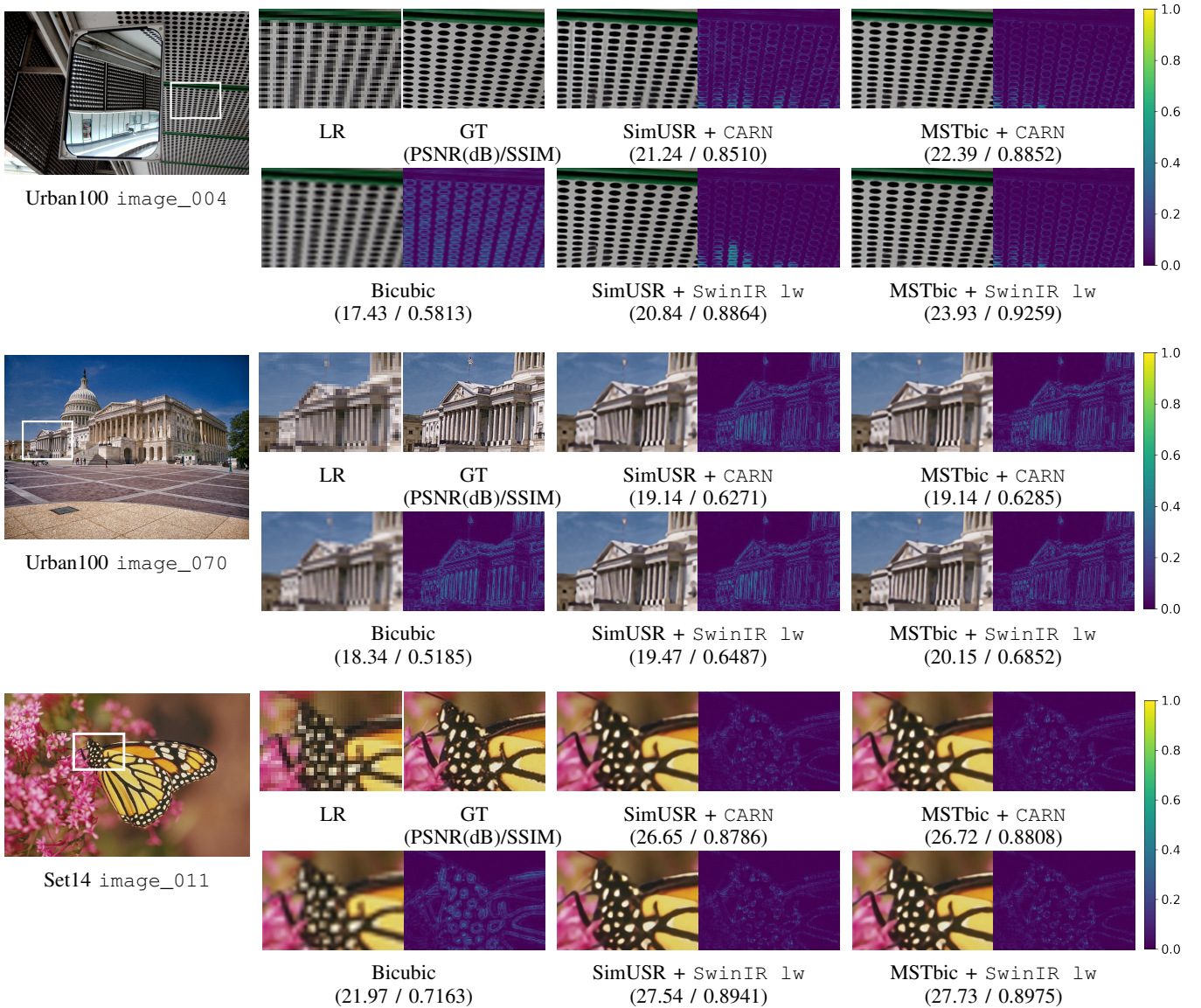

\captionsetup[subfigure]{justification=centering} 
\input{figures/quali_comp/urban100_004_v4_subfigure}
\vfill \vspace{-1mm}
\input{figures/quali_comp/urban100_070_v4_subfigure}
\vfill \vspace{-1mm}
\input{figures/quali_comp/set14_011_v4_subfigure}
\vfill \vspace{-4mm}
\caption{\textbf{Visual comparison} of image patches, which are marked with a white box in the original test images of Urban100 and Set14 on the left. Shown are the low resolution (LR), high-resolution ground truth (GT), and 4x SR results for combinations of SimUSR \cite{ahn_simusr_2020} or MSTbic (ours) training methods with \texttt{CARN} or \texttt{SwinIR lw} models, as well as the result for the bicubic baseline. For each patch, PSNR (dB) and SSIM are calculated and the error map is shown.}
    \label{fig:comparison}
\vspace*{-1mm}
\end{figure*}

Comparing SimUSR and MSTbic side-by-side for the \texttt{CARN} model, MSTbic provides slight improvements or is at least on par across all datasets, with the single exception of Urban100.
For the in any case stronger performing \texttt{SwinIR lw} transformer model, \textit{\mbox{MSTbic} improves performance on all datasets and over all metrics compared to SimUSR, proving to be the strongest model/training method combination}.
Note that the standard deviations show significance in most cases.

Next, we qualitatively compare challenging image patches from Urban100 and Set14 in Fig.\ \ref{fig:comparison}. 
For each result, an error map visually represents the difference between corresponding pixels in the ground truth (GT) and the result image patch, here normalized to a range of 0 to 1.
First we compare the baseline methods bicubic interpolation and SimUSR\cite{ahn_simusr_2020} + \texttt{CARN} with our method MSTbic + \texttt{SwinIR lw}. 
Examining \texttt{image\_070} of Urban100 as an example, the pillars of the building are more detailed and better separated compared to both baselines.
For \texttt{image\_011} of Set14 showing multiple sharply separated colored regions, the bicubic baseline is drastically outperformed, while the difference between SimUSR\cite{ahn_simusr_2020} + \texttt{CARN} and  MSTbic + \texttt{SwinIR lw} is more subtle, yet in \mbox{favor} of the latter.
Comparing \texttt{CARN} and \texttt{SwinIR lw} for all image patches, the most substantial reconstruction errors occur at the boundaries of fine image structures, and most of the time \texttt{SwinIR lw} models seem to perform better in these detailed regions.

\subsection{Ablation Studies}
Here, we show the adaptation of the MST\cite{moller_super-resolution_2023} method to macroscopic LR images created by bicubic degradation to yield our MSTbic training method. Also, we perform ablations on the interpolation kernels for scale augmentation.

\begin{table*}[!htb]
	\centering
	\caption{Adapting multi-scale training (MST) to macroscopic images employing \texttt{SwinIR lw}: MSTbic parameter search for a {\bf 4x bicubic SR task} and DIV2K\textsuperscript{val}. We report PSNR (dB) and SSIM on the Y-channel and also show test performance on benchmark datasets. Models were {\bf trained} from scratch based {\bf on LR images of DF2K\textsuperscript{train}}. Original MST\cite{moller_super-resolution_2023} uses multiple kernels for degradation, while our method MSTbic only uses the bicubic kernel for this task. Best results are in bold font.}
   \resizebox{1\textwidth}{!}{
	\begin{tabular}{lcc|c@{\hskip 4pt}c|c@{\hskip 4pt}cc@{\hskip 4pt}cc@{\hskip 4pt}cc@{\hskip 4pt}cc@{\hskip 4pt}cc@{\hskip 4pt}c}
		\toprule
		  Method& \multicolumn{2}{c|}{Parameters} & \multicolumn{2}{c|}{DIV2K\textsuperscript{val}} & \multicolumn{2}{c}{Set5} & \multicolumn{2}{c}{Set14} & \multicolumn{2}{c}{BSD100} & \multicolumn{2}{c}{Urban100} & \multicolumn{2}{c}{Manga109} \\
            &  $\alpha^{\mathrm{min}}$ & $\beta^{\mathrm{max}}$ & PSNR & SSIM & PSNR & SSIM & PSNR & SSIM & PSNR & SSIM & PSNR & SSIM & PSNR & SSIM \\ \midrule
MST \cite{moller_super-resolution_2023} & $0.25^{*}$ & $2.5^{*}$ & 30.41 &0.8395 &32.10 &0.8946 &28.39 &0.7867 &27.63 &0.7392 &26.10 &0.7892 &30.43 &0.9091   \\
MST\cite{moller_super-resolution_2023} bic & $0.25^{*}$ & $2.5^{*}$ & 30.50 &0.8409 &32.17 &0.8955 &28.46 &0.7868 &27.67 &0.7403 &26.20 &0.7923 &30.65 &0.9129 \\
\midrule
\multirow{4}{*}{MSTbic (ours)}
& 1 & 1     & 30.53 & 0.8415 & 32.23 & 0.8961 & 28.46 & 0.7872 & 27.68 & 0.7409 & 26.25 & 0.7944 & 30.82 & 0.9155 \\
& 0.95 & 1.05 & \textbf{30.60} &  \textbf{0.8425}  &   32.25  &  0.8959    & 28.53 &\textbf{0.7880} &27.70 & 0.7414 & 26.31 & 0.7956 & 31.08 & \textbf{0.9169}  \\
& 0.9 & 1.1  & \textbf{30.60} & \textbf{0.8425} &\textbf{32.27} &\textbf{0.8963} & 28.52 & 0.7874 & \textbf{27.71} & \textbf{0.7415} & \textbf{26.34} & \textbf{0.7962} & 31.07 &\textbf{0.9169} \\
& 0.8 & 1.2 &\textbf{30.60} &0.8424 &32.25 &0.8962 &\textbf{28.54} &0.7878 &\textbf{27.71} &0.7414 &26.30 &0.7957 & \textbf{31.09} & 0.9167   \\
  
  \bottomrule
\end{tabular}
    }
	\label{tab:scale_ablations}
\end{table*}
\newcommand{\wip}[1]{\textcolor{black}{#1}}
\begin{table*}[!htb]
	\centering
	\caption{{\bf Scale augmentation kernel ablations} for MSTbic (ours) employing \texttt{SwinIR lw}: Scale augmentations use $\alpha^\mathrm{min}=0.9$ and $\beta^\mathrm{max}=1.1$. Models were {\bf trained} based {\bf on} {\bf LR images of DF2K\textsuperscript{train}} and tested on the {\bf 4x bicubic SR task} and benchmark datasets. The $\uparrow$ and $\downarrow$ symbols denote the usage of this kernel for downscaling or upscaling, respectively, while the kernels are abbreviated as NN (nearest neighbor), Bic (Bicubic), Bil (Bilinear), Ham (Hamming), Lanc (Lanczos), and Box. * marks MST\cite{moller_super-resolution_2023} bic scaling parameter settings, given for reference. We report PSNR (dB) and SSIM on the Y-channel. Best results are in bold font.}
   \resizebox{1\textwidth}{!}{
	\begin{tabular}{cccccc|c@{\hskip 4pt}c|c@{\hskip 4pt}cc@{\hskip 4pt}cc@{\hskip 4pt}cc@{\hskip 4pt}cc@{\hskip 4pt}cc@{\hskip 4pt}c}
		\toprule
		  \multicolumn{6}{c|}{Scale Augmentation Kernels} & \multicolumn{2}{c|}{DIV2K\textsuperscript{val}} & \multicolumn{2}{c}{Set5} & \multicolumn{2}{c}{Set14} & \multicolumn{2}{c}{BSD100} & \multicolumn{2}{c}{Urban100} & \multicolumn{2}{c}{Manga109} \\
             NN & Bic & Bil & Ham & Lanc & Box  & PSNR & SSIM & PSNR & SSIM & PSNR & SSIM & PSNR & SSIM & PSNR & SSIM & PSNR & SSIM \\ \midrule
$\downarrow^{*}$ & $\uparrow^{*}$ & &&&& 30.50 &0.8409 &32.17 &0.8955 &28.46 &0.7868 &27.67 &0.7403 &26.20 &0.7923 &30.65 &0.9129  \\
\midrule
   &          &   &     &      &    & 30.53 & 0.8415 & 32.23 & 0.8961 & 28.46 & 0.7872 & 27.68 & 0.7409 & 26.25 & 0.7944 & 30.82 & 0.9155 \\
  $\boldsymbol{\downarrow}$ &          &   &     &      &     & \wip{30.52} & \wip{0.8408} & \wip{32.22} & \wip{0.8957} & \wip{28.40} & \wip{0.7862} & \wip{27.68} & \wip{0.7402} & \wip{26.24} & \wip{0.7927} & \wip{30.78} & \wip{0.9133}\\
  $\boldsymbol{\downarrow}$ & $\uparrow$ & &&& \hspace{-1.4cm}(MSTbic, ours)& \textbf{30.60} & \textbf{0.8425} &32.27 &0.8963 & \textbf{28.52} & 0.7874 & \textbf{27.71} & \textbf{0.7415} & 26.34 &0.7962 & 31.07 &0.9169 \\
  $\boldsymbol{\downarrow}$ & $\uparrow$ &  & $\uparrow$ & $\uparrow$ &  &\textbf{30.60} &0.8423 &32.26 &0.8960 &28.49 &0.7878 &\textbf{27.71} &0.7415 &\textbf{26.35} &0.7965 &31.07 & \textbf{0.9173} \\
  $\boldsymbol{\downarrow} \uparrow$ & $\uparrow$  & $\uparrow$ & & & &30.59 &0.8423 &32.27 &0.8962 &28.49 &0.7879 &27.70 &0.7414 & \textbf{26.35} & \textbf{0.7966} &31.06 &0.9169\\
  $\boldsymbol{\downarrow} \uparrow$ & $\uparrow$  & $\uparrow$ & $\uparrow$ & $\uparrow$ & $\uparrow$ & \textbf{30.60} &0.8424 &\textbf{32.30} & \textbf{0.8966} & 28.50 & \textbf{0.7879} & \textbf{27.71} & 0.7414 & \textbf{26.35} & 0.7962 & 31.04 & 0.9170  \\
  \midrule
   & $\boldsymbol{\uparrow}$ &  &  &  & &\wip{30.55}&\wip{0.8409}&\wip{32.20} &\wip{0.8955} &\wip{28.45} &\wip{0.7866} &\wip{27.67} &\wip{0.7401} &\wip{26.19} &\wip{0.7918} &\wip{30.97} &\wip{0.9155} \\
    $\downarrow$ & $\downarrow \boldsymbol{\uparrow}$ & $\downarrow$ & &  & & \textbf{30.60} &0.8423 &32.27 &0.8963 &28.50 &0.7876 &27.70 &0.7412 &26.32 &0.7961 &\textbf{31.13} &0.9172 \\
   \midrule
    $\downarrow \uparrow$ & $\downarrow \uparrow$ & $\downarrow \uparrow$ &  &  & & \textbf{30.60} &0.8423 &32.24 &0.8959 &28.49 &0.7877 &27.70 &0.7413 &26.32 &0.7958 &31.10 &0.9171 \\
    $\downarrow \uparrow$ & $\downarrow \uparrow$ & $\downarrow \uparrow$ & $\downarrow \uparrow$ & $\downarrow \uparrow$ & $\downarrow$ $\uparrow$ & \textbf{30.60} &0.8423 & \textbf{32.30} &0.8965 &28.50 &0.7878 &27.70 &0.7414 &26.34 &0.7961 &31.08 &0.9171\\
  \bottomrule
\end{tabular}
    }
	\label{tab:kernel_ablations}
\end{table*}
\paragraph*{Bicubic degradation} First, we replace the degradation function.
In Table \ref{tab:scale_ablations}, first segment, we replace the set of six different downscaling kernels used in MST\cite{moller_super-resolution_2023} by using the bicubic kernel only (MST\cite{moller_super-resolution_2023} bic). This improves performance on our task as anticipated but remains suboptimal due to the use of the original scaling augmentation parameter values.

\paragraph*{Scale augmentation range}
Next, we search for optimal scaling augmentation parameter values and conduct experiments on the range of the scale augmentation, defined by downscaling parameter $\alpha^{\mathrm{min}}$ and upscaling parameter $\beta^{\mathrm{max}}$, condensed in Table \ref{tab:scale_ablations}, second segment. For this parameter search, we use nearest neighbor (NN) downscaling and bicubic (Bic) upscaling as scale augmentation kernels and thereby adapt original MST to macroscopic images: MSTbic.
We find that the relatively large scaling range ($\alpha^{\mathrm{min}}\!=\!0.25$, $\beta^{\mathrm{max}}\!=\!2.5$) of both MST\cite{moller_super-resolution_2023} from the microscopy domain and MST\cite{moller_super-resolution_2023} with bicubic degradation only is suboptimal for this task, presumably due to the higher level of detail in the macroscopic images.

We select  $\alpha^{\mathrm{min}}\!=\!0.9$, $\beta^{\mathrm{max}}\!=\!1.1$  based on the validation data DIV2K\textsuperscript{val}, while the metrics indicate similar performance for small scaling factors, and show the performance on the various test sets.
We also confirm the effect of scale augmentation in general versus training without scale augmentation ($\alpha^{\mathrm{min}}\!=\!1$, $\beta^{\mathrm{max}}\!=\!1$).

\paragraph*{Scale augmentation kernels}
To investigate the effects of downscaling ($\downarrow$) and upscaling ($\uparrow$) as well as the kernel used, we perform ablation studies on DIV2K\textsuperscript{val} in \mbox{Table \ref{tab:kernel_ablations}}.
Compared to no scale augmentation, we surprisingly find that using only downscaling with nearest neighbor interpolation for scale augmentation decreases performance (2nd table segment).
On the other hand, utilizing only upscaling with the bicubic kernel improves PSNR, yet also decreases SSIM (3rd table segment).
Based on the validation data DIV2K\textsuperscript{val}, the MSTbic configuration (nearest neighbor downscaling and bicubic upscaling) still indicates a marginal advantage over all other configurations (top in SSIM = 0.8425), so we adopt this for our final proposal.
Note that on the benchmark (test) sets other configurations are even slightly better at times.
Also, including additional kernels for downscaling and upscaling yields quite comparable results (last table segment).
For the test data, the MSTbic configuration performs competitive, yet supplementing the bicubic kernel with additional kernels for upscaling can further enhance generalizability for some datasets.

\subsection{Limitations}\label{sec:limitations}
The LR-only benchmark simulates its unsupervised nature by allowing only images for model training, which were downscaled with the bicubic kernel to 25\% of original width and height for the 4x SR task.
These low-passed images appear blurred and are reduced in high-frequency image details.
While the benchmark task involves the restoration of such fine textures and sharpness in test images, the scale augmentation within our MSTbic method does not reintroduce high-frequencies in pseudo-HR training targets. (This equally holds for SimUSR.)
In fact, the bicubic kernel for upscale augmentation further removes some high-frequency information.
Therefore, MSTbic is somehow limited by the quality of LR images, from which training targets are created, yet succeeds by increasing the variety of the training data.
It has furthermore been demonstrated in SR that certain augmentations, which significantly alter the neighborhood relationships of pixels, can negatively impact SR performance \cite{yoo_rethinking_2020}. For our MSTbic training method we also show that too high scaling augmentation factors are harmful for SISR performance, particularly when applied to detailed photographs of urban environments and real-world scenes included in the benchmark datasets.
\section{Conclusions}\label{sec:conc}
In this work, we are the first to deploy a \textit{transformer}-based super-resolution model to the LR-only single-image super-resolution benchmark, where high-resolution images are not available for training, yet used for evaluation.
For this unsupervised task, we also propose an LR-only training method MSTbic, that we adapt from microscopy super-resolution to the macroscopic benchmark image domain.
MSTbic creates pseudo-LR/HR training pairs based on the available low-resolution training images and leverages {\it both} image downscaling and upscaling for image scale augmentation.
Thereby, we show improved results for a lightweight convolutional neural network model, but particularly with and for a lightweight transformer model, thereby setting a new state-of-the-art performance for this LR-only benchmark on both network types.

{
\bibliographystyle{latex/ieee_fullname} 
\bibliography{main.bbl}
}
\clearpage

\end{document}